\documentclass[aps,prd,showpacs,superscriptaddress,twocolumn
,amsmath,amssymb]{revtex4}
\usepackage{graphicx}
\usepackage{dcolumn}
\usepackage{bm}
\usepackage{subfigure}
\usepackage{float}

\def\agt{\mathrel{\raise.3ex\hbox{$>$}\mkern-14mu\lower0.6ex\hbox{$\sim$}}}
\def\alt{\mathrel{\raise.3ex\hbox{$<$}\mkern-14mu\lower0.6ex\hbox{$\sim$}}}

\def\beq{\begin{equation}}
\def\eeq{\end{equation}}
\def\bsubeq{\begin{subequations}}
\def\esubeq{\end{subequations}}
\def\Re{{\rm Re}}
\def\Im{{\rm Im}}
\def\pd{\partial}
\def\cc{{\rm c.c.}}
\def\Ei{{\rm Ei}}
\def\supp{{\rm supp}}

\begin{document}

\title{Possible discovery of a nonlinear tail and second-order quasinormal modes \\ in black hole ringdown}

\author{Satoshi Okuzumi}
\affiliation{
Graduate School of Human and Environmental Studies, Kyoto University, Kyoto,
 606-8501, Japan
}
\author{Kunihito Ioka}
\affiliation{Theory Division, KEK (High Energy Accelerator Research Organization), 1-1 Oho, Tsukuba 305-0801, Japan}
\author{Masa-aki Sakagami}
\affiliation{
Graduate School of Human and Environmental Studies, Kyoto University, Kyoto,
 606-8501, Japan
}

\begin{abstract}
We investigate the nonlinear evolution of black hole ringdown in the framework of higher-order metric perturbation theory.
By solving the initial-value problem of a simplified nonlinear field model analytically as well as numerically,
we find that (i) second-order quasinormal modes (QNMs) are indeed excited at frequencies different from those of first-order QNMs, as predicted recently.
We also find serendipitously that (ii) the evolution is dominated by a new type of power-law tail at late times.
This ``second-order power-law tail'' decays more slowly than any late-time tails known in the first-order (i.e., linear) perturbation theory, and is generated at the wavefront of the first-order perturbation by an essentially nonlinear mechanism.
These nonlinear components should be particularly significant for binary black hole coalescences, and 
could open a new precision science in gravitational wave studies. 
\end{abstract}

\maketitle

\section{Introduction}
Direct detection of gravitational waves is one of the most exciting challenges in astrophysics today.
It will not only enable to verify general relativity, 
but will also provide a new observational window 
toward the universe. 
Current and future gravitational wave detectors such as LIGO, LISA,
 and DECIGO/BBO should make it a reality in the near future.
 
The ``ringdown'' of black holes is an important target of gravitational-wave observations. 
It is known theoretically that black holes have characteristic oscillation modes, called {\it quasinormal modes}
(QNMs) \cite{Nollert99,Kokkotas99}.
QNMs are the solutions to perturbation equations which satisfy so-called ``outgoing boundary conditions,''
i.e., which propagate purely inward at the horizon and purely outward at spatial infinity.    
They are distinguished from ordinary normal modes because they decay at certain rates, having complex frequencies.
The remarkable property of the black hole QNMs is that their frequencies are uniquely determined by the mass and spin of black holes.
This means that the parameters of black holes can be directly measured by observing their ringdown through the gravitational waves.

The black hole ringdown is particularly prominent in binary black hole mergers \cite{FlanaganHughes98}. 
Recently, numerical relativity has succeeded in calculating their entire evolution \cite{Pretorius05,Campanelli+06,Baker+06}.
As a result, it has been found that the least-damped QNM dominates the gravitational waves radiated during the ringdown phase, carrying away $\sim 1\%$ of the initial rest mass energy of the binary system \cite{Berti+07}. 
The ringdown wave with such large energy will be detected by future gravitational-wave detectors with a
high signal-to-noise ratio \cite{FlanaganHughes98}.

Usually, the black hole ringdown is understood as the response of a black hole to {\it linear} perturbations \cite{Leaver86,SunPrice88,SunPrice90,NollertSchmidt92,Andersson97}.
In black hole perturbation theory, the metric is divided into a stationary black hole background (formed after a binary merger, for instance) and small perturbations on it.
Now let us consider an {\it initial-value} problem \cite{Leaver86} where the perturbations are initially
distributed near the horizon.
Then, the linear theory predicts that an observer far away from the black hole sees two different 
stages in the evolution of gravitational waves.
At early times, the evolution is dominated by quasinormal (QN) ringing, a superposition of QNMs.
At late times, in contrast,  the evolution is characterized by a power-law tail, which is related to the backscattering of the QN ringing far away from the black hole \cite{Ching95}.
Also, the initial-value analysis clarifies another important fact that the QNMs are excited only around a black hole \cite{Andersson97}.
As a result, evolved QNMs are always {\it truncated} by a future-directed light cone \cite{Szpak04}.
This truncation is essential for QNMs, since without it they would diverge at both spatial infinity and the horizon. 

Although the linear theory is clearly useful,
the recent numerical simulations imply that the {\it nonlinearity} of the ringdown is also worth studying.
The radiated energy of $\sim 1\% \times {\rm (rest~mass~energy)}$ is translated into the amplitude $\sim 10\%$  of the first-order (linear) metric perturbations,
implying that the next-leading, second-order ones will have amplitude as large as $\sim 10\%$ of the first-order amplitude.
This strongly suggests that the second- and higher-order perturbations should contribute the ringdown waveform to that extent, providing us with some additional information on the merger events and the nonlinearity of general relativity.

Despite its potential importance, the nonlinearity of black hole ringdown has been hardly investigated.
Only recently, one of the authors (K. I.) and his coworker have for the first time applied second-order metric perturbation theory to studying QNMs of Schwarzschild black holes \cite{IokaNakano07,NakanoIoka07}.
Because of the nonlinearity of the Einstein equation, two normal modes in first-order perturbations [say $\cos(\Omega_1t)$ and $\cos(\Omega_2t)$] couple each other, yielding their ``sum tone'' $\propto \cos[(\Omega_1+\Omega_2)t]$ and ``difference tone'' $\propto \cos[(\Omega_1-\Omega_2)t]$ in second-order ones.
The authors found that the sum tones and difference tones of the first-order QNMs also satisfy the outgoing boundary conditions, meaning that they are the QNMs {\it in the second-order perturbation theory}.
They concluded that if these ``second-order QNMs'' are excited in a binary black hole merger, they will have enough energy to be detected by future gravitational-wave detectors.
Detection of the second-order QNMs has many possible applications, such as measuring the distance from the source system, testing the nonlinearity of general relativity, and rejecting fake events in ringdown searches.  

Questions have remained unanswered.
First, are the second-order QNMs really excited 
even if the first-order ones are {\it spatially truncated}?
The previous work \cite{IokaNakano07,NakanoIoka07} was done in the {\it frequency domain},
where the first-order QNMs (or the sources of the second-order perturbations)
 are implicitly assumed to be extended from the horizon to spatial infinity.
This assumption is actually invalid, because the sources would diverge far away from the black hole.
The first-order QNMs are actually truncated at a light cone, and thus 
the sources of the second-order QNMs are distributed in a {\it time-dependent} way.
Second, does the second-order theory provide {\it only} the second-order QNMs?
In general, first-order QNMs are accompanied by a power-law tail,
as naturally shown using the {\it time-domain} Green's function of the perturbations
\cite{Leaver86,Ching95,Andersson97}.
On the other hand, it has been open whether the second-order QNMs are accompanied 
by any other wave components, especially {\it of nonlinear origin}.
To address these questions, one must move away from the frequency-domain analysis 
and treat the evolution of the second-order perturbations as a {\it time-dependent problem}.

In this study, we give answers to the above questions 
by solving an initial-value problem for nonlinear black hole ringdown. 
Our analytic calculation is based on a time-domain analysis developed
for linear black hole perturbations \cite{Leaver86,Andersson97}.
As far as we know, this analysis has been never applied to nonlinear black hole perturbations.
Also, we employ a simple nonlinear field model having key properties 
in common with black hole perturbations.
Our nonlinear model is attractive, 
since the original equations for black hole perturbations are too complicated to be treated analytically, and 
since direct simulations using numerical relativity would not be accurate enough 
to resolve nonlinear evolution in ringdown phases at the present time \cite{IokaNakano07}.
Our model enables us not only to solve the problem analytically,
but also to confirm the results with fully nonlinear simulations in a good accuracy.

Our answers to the above questions are as follows. First, the second-order QNMs are indeed excited by {\it truncated} first-order QNMs. 
Second, the second-order QNMs are accompanied by a  power-law tail,
 but it is {\it essentially different} from the tails in the linear perturbation theory \cite{Ching95}.
This tail, to which we shall refer as ``second-order power-law tail,'' decays more slowly than any of the first-order tails, and even dominates the {\it fully nonlinear} evolution of the model field at late times.
This means a surprising fact that the first-order theory {\it fails} to predict the late-time behavior of the ringdown.
We also find that the behavior of the second-order tail is completely determined
by the spatial dependence of an effective ``source'' term for second-order perturbations.
Since the effective source term of our model behaves in essentially the same manner
 as that of black hole perturbations,
we strongly expect that the second-order power-law tail should appear in real black hole ringdown,
and should play a leading role in the late-time evolution.

This paper is organized as follows. In Sec.~II, we introduce a model system for black hole perturbations 
and derive the perturbation equations used in the subsequent sections.
In Sec.~III, we analytically solve an initial-value problem of the perturbations with an approximated Green's function, and decompose the evolved second-order perturbation into different components.
In Sec.~IV, we confirm the validity of the analytic calculations by numerical simulations. 
In Sec.~V, we present our conclusion and discuss several possible applications to future gravitational-wave studies.
We also discuss briefly a relation between our second-order power-law tail and nonlinear tails recently found in other systems.
In Appendix A, we summarize how to construct the time-dependent Green's function using Andersson's ``asymptotic approximation'' \cite{Andersson97}.

\section{Nonlinear field model}
Before introducing our nonlinear field model, 
we briefly review how the equations of black hole perturbations are 
derived from the Einstein equation. The Einstein equation in vacuum reads
\beq
{\bm G}[{\bm g}] = 0, \label{eq:Einstein}
\eeq
where ${\bm G}$ is the Einstein tensor and ${\bm g}$ is a metric tensor. 
In metric perturbation theory, the metric ${\bm g}$ is expanded 
into a stationary background part ${\bm g}^{(0)}$ and perturbation parts ${\bm h}^{(j)}$ ($j=1,2,\cdots$):
\beq
{\bm g} = {\bm g}^{(0)} + \epsilon {\bm h}^{(1)} + \epsilon^2 {\bm h}^{(2)} + \cdots,
\eeq
where $\epsilon$ is an expansion parameter.
The wave equations for ${\bm h}^{(j)}$ can be obtained by expanding the Einstein equation \eqref{eq:Einstein} perturbatively with $\epsilon$:
\begin{align}
{\bm G}^{(1)}[{\bm h}^{(1)}] &= 0, \label{eq:Einstein1}\\
{\bm G}^{(1)}[{\bm h}^{(2)}] &= -{\bm G}^{(2)}[{\bm h}^{(1)},{\bm h}^{(1)}], \label{eq:Einstein2}\\
\vdots \notag
\end{align}
where ${\bm G}^{(1)}$ is the linearized Einstein tensor, and ${\bm G}^{(2)}[{\bm a},{\bm b}]$ is a nonlinear part of ${\bm G}$ and is quadratic in ${\bm a}$ and ${\bm b}$. 
On the Schwarzschild background, the perturbations ${\bm h}^{(j)}$
can be expanded into tensor harmonic components $h^{(j)}_{\ell m}$, and Eqs.~\eqref{eq:Einstein1} and \eqref{eq:Einstein2} can be reduced to (1+1)-dimensional master equations \cite{RW57,Zerilli70,Gleiser96CQG,Gleiser96PRL,Gleiser00}
\begin{align}
\Bigl[ -\pd_t^2 + \pd_{r_*}^2 - V_\ell(r) \Bigr]\psi^{(1)}_{\ell m}(t,r) &= 0, 
\label{eq:Sch1} \\
\Bigl[ -\pd_t^2 + \pd_{r_*}^2 - V_\ell(r) \Bigr]\chi^{(2)}_{\ell m}(t,r) &= S^{(2)}_{\ell m}(t,r),
\label{eq:Sch2}
\end{align}
where $\psi^{(1)}_{\ell m}$ and $\chi^{(2)}_{\ell m}$ are, respectively, the master functions for $h_{\ell m}^{(1)}$ and $ h_{\ell m}^{(2)}$ (for their definitions, see \cite{IokaNakano07,NakanoIoka07});
$r_*$ is the tortoise coordinate \cite{RW57}, covering from the horizon ($r_*\to-\infty$) to spatial infinity ($r_*\to\infty$); $V_\ell(r)$ is an effective potential term that has a peak outside the horizon and vanishes as $r_*\to\pm\infty$ (Regge-Wheeler \cite{RW57} or Zerilli \cite{Zerilli70} potential);
and $S^{(2)}_{\ell m}$ is an effective source term, originating from $-\bm{G}^{(2)}[{\bm h}^{(1)},{\bm h}^{(1)}]$ in Eq.~\eqref{eq:Einstein2} and essentially quadratic in $\psi^{(1)}_{\ell m}$.
Note that the left-hand sides of Eqs.~\eqref{eq:Sch1} and \eqref{eq:Sch2} are described by the same linear differential operator $-\pd_t^2 + \pd_{r_*}^2 - V_\ell$, because they both originate from the linearized Einstein tensor ${\bm G}^{(1)}$.

What we want to do in this study is to clarify what types of wave components are created as
the second-order perturbations during black hole ringdown.
To decompose the perturbations into different types of components,
we have to solve the evolution as analytically as possible. 
However, it would be challenging to do this by a direct use of Eq.~\eqref{eq:Sch2}, 
since the source term $S^{(2)}_{\ell m}$ contains many linearly independent components, each of which depends on $r_*$ in a complicated way (see, e.g., Eq.~(13) in Ref.~\cite{IokaNakano07}).
In addition, it would be also challenging to confirm the results with full-order computations of the Einstein equation \eqref{eq:Einstein}, since current numerical relativity may not be accurate enough to resolve the evolution of the second-order perturbations \cite{IokaNakano07,NakanoIoka07}.
Therefore, we prefer for our purpose to consider a simple model instead of the real perturbations.
  
\begin{figure}
\centering
\includegraphics[width=8cm]{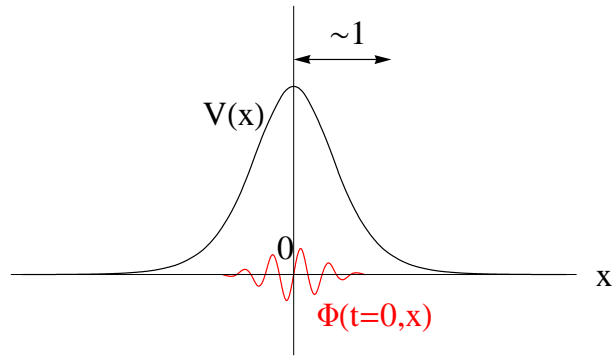}
\caption{Schematic illustration of the potential barrier $V(x)$ and the initial data $\Phi(t=0,x)$. The potential barrier vanishes at $|x|\gg1$, and the initial data has a
compact support near the potential peak located at $x=0$.}
\label{fig:initial}
\end{figure}

Now we introduce our model.   
To avoid oversimplification, 
we require the model to satisfy that 
(i) its first- and second-order
perturbations obey (1+1)-dimensional
wave equations with a potential barrier,
like Eqs.~\eqref{eq:Sch1} and \eqref{eq:Sch2},
and that (ii) the effective source term for the second-order perturbation
 is quadratic in the first-order one.
As one of the simplest models, we adopt a nonlinear, 
(1+1)-dimensional scalar field obeying
\beq
\Bigl[ -\pd_t^2 + \pd_x^2 - V(x) \Bigr]\Phi(t,x) = F(x)\Phi(t,x)^2,
\label{eq:wave0}
\eeq
where $V(x)$ is a potential term and $F(x)$ is a function specified later.
We assume that $V(x)$ has a peak at $x=0$ and vanishes at $|x|\gg1$, as schematically shown in Fig.~\ref{fig:initial}.
Since $x$ corresponds to the tortoise coordinate $r_*$ for black holes,
we refer to the limits $x\to-\infty$ and $x\to+\infty$ as ``horizon'' and ``infinity,'' respectively.
We also assume an observer's position near infinity, $x\gg1$.

It is easily checked that this model meets the above requirements.
We expand $\Phi(t,x)$ in terms of an expansion parameter $\epsilon$ to write
\beq
\Phi(t,x) = \epsilon\phi^{(1)}(t,x) + \epsilon^2\phi^{(2)}(t,x) + \cdots.
\label{eq:pert}
\eeq 
Substituting this into Eq.~\eqref{eq:wave0}, we obtain an infinite set of perturbation equations,
the first two of which read
\begin{align}
\Bigl[ -\pd_t^2 + \pd_x^2 - V(x) \Bigr]\phi^{(1)} &= 0, 
\label{eq:wave1} \\
\Bigl[ -\pd_t^2 + \pd_x^2 - V(x) \Bigr]\phi^{(2)} &= S^{(2)}(t,x),
\label{eq:wave2}
\end{align} 
where 
\beq
S^{(2)}(t,x) \equiv F(x)\bigl(\phi^{(1)}\bigr)^2
 \label{eq:S_def}
\eeq
is an effective source term for the second-order perturbation $\phi^{(2)}$.
It is clear from Eqs.~\eqref{eq:wave1}--\eqref{eq:S_def} that this model does satisfy the requirements (i) and (ii).  

In this study, we consider an initial-value problem of Eq.~\eqref{eq:wave0}
with the following initial conditions:
\begin{gather}
\Phi(t=0,x) = f(x), \label{eq:ICf}\\
\pd_t\Phi(t=0,x) = g(x), \label{eq:ICg}
\end{gather}
where $f(x)$ and $g(x)$ are arbitrary functions vanishing far away from the potential peak
($|x|\gg$1), as schematically illustrated in Fig.~\ref{fig:initial}.
To solve the problem perturbatively, we have to divide these initial data among the perturbation quantities $\phi^{(j)}$.
Since there is no unique choice of dividing the data, 
we adopt the simplest choice such that the first-order perturbation theory is exact {\it at the initial time}:
\begin{gather}
\phi^{(1)}(t=0,x) = f(x), \label{eq:IC1}\\
\pd_t\phi^{(1)}(t=0,x) = g(x), \\
\phi^{(k)}(t=0,x) = 0, \\
\pd_t\phi^{(k)}(t=0,x) = 0, \label{eq:IC2}
\end{gather}
where $k\geq2$. 
Note that this choice does {\it not} guarantee that the first-order theory 
remains to be a good approximation to the full-order theory {\it at all times}.
In fact, as we will see later, $\phi^{(2)}$ {\it does}
surpass $\phi^{(1)}$ in amplitude, dominating the late-time behavior of $\Phi$. 

\section{Analytic calculations}
Since all the perturbations $\phi^{(j)}\,(j=1,2,\cdots)$ obey linear equations with the same operator $-\pd_t^2 + \pd_x^2 - V(x)$, 
their evolution is determined by a single Green's function.
The retarded Green's function $G(\tau;x,x')$ for $\phi^{(j)}$ is the solution to
\beq
\Bigl[ -\pd_t^2 + \pd_x^2 - V(x) \Bigr]G(t-t';x,x') = \delta(t-t')\delta(x-x'),
\label{eq:waveG}
\eeq
under the retarded boundary condition 
\beq
G(\tau;x,x')=0, \quad \tau<0.
\eeq
Using this Green's function,
the solutions to Eqs.~\eqref{eq:wave1} and \eqref{eq:wave2} with the initial conditions \eqref{eq:IC1}--\eqref{eq:IC2} are written as
\begin{align}
\phi^{(1)}(t,x) &= \iint_{-\infty}^\infty dt'dx'  G(t-t';x,x')I(t',x'),
\label{eq:phi1_sol} \\
\phi^{(2)}(t,x) &= \iint_{-\infty}^\infty dt'dx'  G(t-t';x,x')S^{(2)}(t',x'),
\label{eq:phi2_sol} 
\end{align}
where $I(t,x) = -f(x)\delta'(t) -g(x)\delta(t)$.

If the background were ``flat'' ($V\equiv0$), Eq.~\eqref{eq:waveG} would be easily solved to give
\beq
G(t-t';x,x') = -\frac{1}{2}\Theta(t-t'-|x-x'|),
\label{eq:G_flat}
\eeq
where $\Theta(z)$ is the step function.
Equation~\eqref{eq:G_flat} means that an observer at $(t,x)$ receives waves generated inside a past-directed light cone, $t' \leq t-|x-x'|$. 
Also, the ``flatness'' of the Green's function for $t' \leq t-|x-x'|$ represents that the waves would propagate without being reflected or deformed.
In our case, however, the background is ``curved'' $(V\not\equiv0)$, and some of the waves 
are reflected or deformed by the potential barrier.
For this reason, the Green's function is much more complicated than Eq.~\eqref{eq:G_flat} and is conventionally divided
into three parts \cite{Leaver86,Andersson97}:
\beq
G(\tau;x,x') = G_F(\tau;x,x')  + G_Q(\tau;x,x') + G_B(\tau;x,x').
\label{eq:G}
\eeq
The ``flat part'' $G_F$ represents the direct propagation of waves in asymptotically 
flat regions $(|x'|\gg1)$.
The ``QNM part'' $G_Q$  is related to the excitation of 
first-order QNMs.
The ``branch-cut part'' $Q_B$ describes the backscattering of waves
by the long-range part of the potential barrier.

Unfortunately, the exact expression of the Green's function~\eqref{eq:G} is unknown for most cases including black hole perturbations. Hence, we construct an approximate Green's function with Andersson's ``asymptotic approximation'' \cite{Andersson97}.
Using this approximation and assuming an observer's position at infinity ($x\gg1$), 
the flat part $G_F$ and the QNM part $G_Q$ are obtained as (see Appendix A)
\begin{align}
G_F(\tau;x,x') &\approx -\frac{1}{2}\Theta(\tau-|x-x'|)\Theta(x+x'-\tau), \label{eq:G_F}\\
G_Q(\tau;x,x') &\approx -\frac{1}{2}\Theta(\tau-x-|x'|) \notag \\
&\quad\times\sum_{n=0}^{\infty} 
\biggl[\frac{e^{s_n(\tau-x)}}{s_nA_L'(s_n)}y_n(x') +\cc \biggr], \label{eq:G_Q}
\end{align}
where $\{y_n\}_{n=0}^\infty$ are the first-order QNMs defined as the homogeneous solutions to the Laplace transform of Eq.~\eqref{eq:waveG} with the ``outgoing boundary conditions''
\beq
y_n(x) \approx \begin{cases}
e^{s_n x}, & x\to-\infty, \\
B_L(s_n)e^{-s_n x}, & x\to+\infty.
\end{cases}
\label{eq:QNM}
\eeq
Note that  ``QN frequencies'' $\omega_n$, often used in the literature,
are related to $s_n$ by $\omega_n = is_n$.
It is also noted that the branch-cut part $G_B$ is not derived from the asymptotic approximation,
because this approximation implicitly neglects long-range components of the potential barrier \cite{Andersson97}.
However, the contributions of $G_B$ (i.e., late-time tails in the first-order theory) is usually small compared to the other contributions
(see, e.g., \cite{Ching95}), and hence we neglect $G_B$ in the following analysis. 

\begin{figure*}[t]
\centering
\includegraphics[width=18cm]{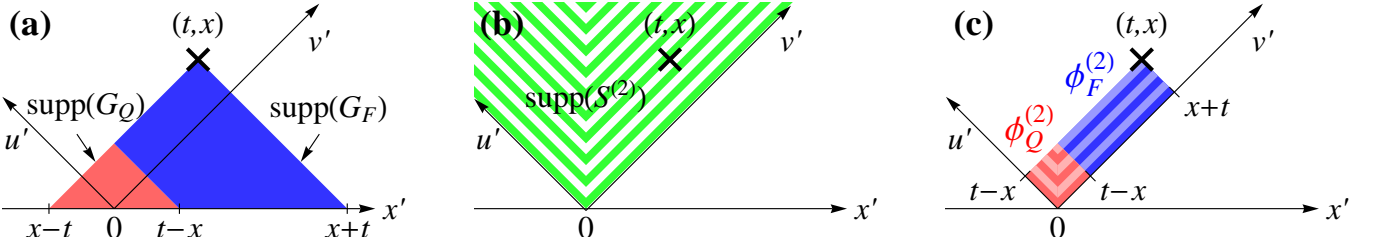}
\caption{(a) Space-time diagram describing the support of the Green's function $G(t-t';x,x')$ 
[Eq.~\eqref{eq:G}] for an observer at $x\,(<t)$. 
$\supp(G_F)$ and $\supp(G_Q)$ represent the supports of the flat part $G_F$ [Eq.~\eqref{eq:G_F}] and
QNM part $G_Q$ [Eq.~\eqref{eq:G_Q}], respectively.
$u'=t'-x'$ and $v'=t'+x'$ are the retarded and the advanced null coordinates. 
(b) Support of the source term $S^{(2)}(t',x')$ [Eq.~\eqref{eq:S}], which is equal to that of the evolved first-order perturbation $\phi^{(1)}(t',x')$ [Eq.~\eqref{eq:phi1_asympt}]. The stripes indicate the wave pattern of $\phi^{(1)}$ schematically. 
(c) Integration regions for $\phi^{(2)}_F$ and $\phi^{(2)}_Q$  in Eq.~\eqref{eq:phi2_sol2}, $\supp(G_F)\cap\supp(S^{(2)})$ and $\supp(G_Q)\cap\supp(S^{(2)})$, respectively.}
\label{fig:spacetime}
\end{figure*}
The supports of $G_F(t-t';x,x')$ and $G_Q(t-t';x,x')$ are indicated in Fig.~\ref{fig:spacetime}(a).
For $t'\geq t-x-|x'|$, the Green's function $G\approx G_F$ 
behaves in essentially the same way as that of the flat background, or Eq.~\eqref{eq:G_flat}.
This means that waves generated in this region arrive at $(t,x)$ 
without passing through the potential barrier, and thus without being reflected or deformed.
For $t'\leq t-x-|x'|$, on the other hand, the Green's function 
$G\approx G_Q$ behaves differently from Eq.~\eqref{eq:G_flat}, oscillating with QN frequencies. 
This implies that waves generated in this region pass through the potential, 
and are then deformed into the first-order QNMs.
These facts also imply that the first-order QNMs are excited only around the potential peak,
and are thus always {\it truncated} by a future-directed light cone \cite{Szpak04}.

Now we perform the integration in Eq.~\eqref{eq:phi1_sol} using the approximate Green's function, or Eqs.~\eqref{eq:G}--\eqref{eq:G_Q} and $G_B\approx0$.
Given that $f(x)$ and $g(x)$ are localized near the potential peak,  
the evolution of $\phi^{(1)}$ is calculated to be
\begin{align}
\phi^{(1)}(t,x) &\approx -\Theta(t-x)\int dx'\Bigl[f(x')\pd_t+g(x')\Bigr]G_Q(t;x,x') \notag \\
&\equiv \Theta(t-x)\sum_{n=0}^{\infty}\Bigl[\,C_n e^{s_n(t-x)} + C^*_n e^{s_n^*(t-x)} \,\Bigr] ,
\label{eq:phi1_asympt}
\end{align}
where 
\beq
C_n \equiv \ \frac{1}{2}\int dx'\Bigl[f(x')s_n+g(x')\Bigr]\frac{y_n(x')}{s_nA_L'(s_n)} 
\eeq
is the ``excitation coefficient'' for the first-order QNM $e^{s_n(t-x)}$.
Equation~\eqref{eq:phi1_asympt} means that the initial first-order perturbation evolves into a superposition of the first-order QNMs truncated at $x=t$.
Performing a similar calculation for $x<0$ shows that the first-order QNMs are also truncated at $x=-t$.
The evolution of $\phi^{(1)}$ is schematically illustrated in Fig.~\ref{fig:spacetime}(b).

The evolution of $\phi^{(2)}$ is obtained by substituting
Eq.~\eqref{eq:phi1_asympt} into Eq.~\eqref{eq:S_def}. 
Since $\phi^{(1)}$ vanishes at $x>t$ and $x<-t$,
we can rewrite Eq.~\eqref{eq:phi2_sol} as
\begin{align}
\phi^{(2)}(t,x) &\approx \frac{1}{2}\int_{0}^{\infty}du'\int_{0}^{\infty}dv'\,G_F(t-t';x,x')S^{(2)}(t',x') \notag \\
  &~ + \frac{1}{2}\int_{0}^{\infty}du'\int_{0}^{\infty}dv'\,G_Q(t-t';x,x')S^{(2)}(t',x'), \notag \\
&\equiv \phi^{(2)}_F(t,x) + \phi^{(2)}_Q(t,x),
\label{eq:phi2_sol2}
\end{align}
where we have introduced the retarded and advanced null coordinates, $u\equiv t-x$ and $v\equiv t+x$,
and used the relation $dt'dx' = (1/2)du'dv'$.
$\phi^{(2)}_F(t,x)$ and $\phi^{(2)}_Q(t,x)$ denote the contributions from the flat part and QNM part,
respectively. Their integration regions at $t>x$ are illustrated in Fig.~\ref{fig:spacetime}(c).
 
To carry out the integrations in Eq.~\eqref{eq:phi2_sol2}, 
we must specify the form of the source term $S^{(2)}(t,x)$, or $F(x)$.
Since the (regularized) source term for black hole perturbations [$S^{(2)}_{\ell m}(t,r_*)$ in Eq.~\eqref{eq:Sch2}]
behaves as $\propto r_*^{-2}$ at infinity  \cite{IokaNakano07,NakanoIoka07},
we choose $F(x)$ so that $S^{(2)}(t,x)$ behaves as $\propto x^{-2}$ at $x\to\infty$.
As one of the simplest form, we assume $F(x) = (|x|+1)^{-2}$, or
\beq
S^{(2)}(t,x) = \frac{\phi^{(1)}(t,x)^2}{(|x|+1)^2},
\label{eq:S}
\eeq
which behaves as $\propto x^{-2}$ at $x\to\pm\infty$ \cite{note1}.
Substituting Eqs.~\eqref{eq:phi1_asympt} and \eqref{eq:S} into Eq.~\eqref{eq:phi2_sol2}, 
we have $\phi^{(2)}(t,x) =0$ for $u<0$, and
\begin{align}
\phi_F^{(2)}(t,x)
&\approx  -\frac{1}{4}\int_{0}^u du'\int_u^v dv'\,\frac{1}{(x'+1)^2} \notag \\
&\quad\times
\sum_{n,n'=0}^\infty\Bigl[\,C_nC_{n'}e^{(s_n+s_{n'}) u'} \notag \\
&\qquad\qquad\quad + C_nC^*_{n'}e^{(s_n+s^*_{n'})u'} +\cc \,\Bigr] 
\label{eq:phi2F}
\end{align}
for $u\geq0$. 
This integration is formally performed by introducing an analytic function
\beq
T(z) \equiv e^{z}\,\Ei(-z),
\eeq
where
\beq
\Ei(z) \equiv  -\int_{-z}^\infty \frac{e^{-w}}{w}dw
\eeq
is the exponential integral. Using $T(z)$, we find a useful equality
\begin{align}
&-\frac{1}{4}\int_0^u du'\int_u^v dv'\frac{e^{su'}}{(x'+1)^2} \notag \\
&\quad= -e^{su'}\Bigl\{T\bigl(s(v-u'+2)\bigr)-T\bigl(s(u-u'+2)\bigr)\Bigr\}\biggr|_{u'=0}^{u'=u}.
\end{align}
Hence, we can rewrite Eq.~\eqref{eq:phi2F} as
\begin{align}
\phi^{(2)}_F(t,x) \approx \phi^{(2)}_{2Q}(t,x) + \phi^{(2)}_{T}(t,x),
\label{eq:phi2F_sol}
\end{align}
where $\phi^{(2)}_{2Q}$ and $\phi^{(2)}_{T}$ are the contributions from the interior ($u'=u$) and the edge
($u'=0$) of the source distribution [see Fig.~\ref{fig:spacetime}(c)]:
\begin{widetext}
\begin{align}
\phi^{(2)}_{2Q}(t,x) &\equiv -\sum_{n,n'}\biggl[\;
C_nC_{n'}e^{(s_n+s_{n'})u'}\Bigl\{T\bigl((s_n+s_{n'})(v-u'+2)\bigr)-T\bigl((s_n+s_{n'})(u-u'+2)\bigr)\Bigr\}
+ \cdots \;\biggr]_{u'=u} \notag \\
&= -\sum_{n,n'}\biggl[\;
C_nC_{n'}e^{(s_n+s_{n'})u}\Bigl\{T\bigl(2(s_n+s_{n'})(x+1)\bigr)-T\bigl(2(s_n+s_{n'})\bigr)\Bigr\} 
+ \cdots
\;\biggr],
\label{eq:phi2Q}
\end{align}
\begin{align}
\phi^{(2)}_{T}(t,x) &\equiv +\sum_{n,n'}\biggl[\;
C_nC_{n'}e^{(s_n+s_{n'})u'}\Bigl\{T\bigl((s_n+s_{n'})(v-u'+2)\bigr)-T\bigl((s_n+s_{n'})(u-u'+2)\bigr)\Bigr\} 
+\cdots
\;\biggr]_{u'=0} \notag \\
&= +\sum_{n,n'}\biggl[\;
C_nC_{n'}\Bigl\{T\bigl((s_n+s_{n'})(v+2)\bigr)-T\bigl((s_n+s_{n'})(u+2)\bigr)\Bigr\}
+\cdots 
\;\biggr],
\label{eq:phi2T}
\end{align}
\end{widetext}
where ``$\cdots$'' represents the terms proportional to $C_nC^*_{n'}$ or $C^*_nC^*_{n'}$.
The behavior of these terms is essentially the same as that of the $C_nC_{n'}$ terms.

We are particularly interested in how $\phi^{(2)}_{2Q}$ and $\phi^{(2)}_{T}$ behave
 at the observer's position, or at $x\gg1$.
First, we examine the behavior of $\phi^{(2)}_{2Q}$.
Using the asymptotic expansion of $T(z)$ at $|z|\gg 1$,
\beq
T(z) = \sum_{k=1}^{\infty}(-1)^k\frac{(k-1)!}{z^k} \approx -\frac{1}{z},
\label{eq:T_asympt}
\eeq
we have 
\begin{align}
\phi^{(2)}_{2Q}(t,x) 
&\approx \sum_{n,n'}\biggl[
C_{nn'}e^{(s_n+s_{n'})u} +D_{nn'}e^{(s_n+s^*_{n'})u}+\cc \biggr] \notag \\
\label{eq:phi2Q_inf}
\end{align}
for $x\gg 1$. Here we have defined $C_{nn'}$ and $D_{nn'}$ as
\begin{align}
C_{nn'} &\equiv C_nC_{n'}T\bigl(2(s_n+s_{n'})\bigr), \label{eq:C}\\
D_{nn'} &\equiv C_nC^*_{n'}T\bigl(2(s_n+s^*_{n'})\bigr), \label{eq:D}
\end{align}
and neglected the terms of $O(x^{-1})$.
It is helpful to rewrite Eq.~\eqref{eq:phi2Q_inf} as
\begin{align}
\phi^{(2)}_{2Q}(t,x) 
&\approx \sum_{n,n'} \biggl[ e^{(\gamma_n+\gamma_{n'})u} \Bigl\{
C_{nn'}e^{-i(\Omega_n+\Omega_{n'})u} \notag \\
&\qquad +D_{nn'}e^{-i(\Omega_n-\Omega_{n'})u} +\cc 
\Bigr\} \biggr],
\label{eq:phi2Q_inf2}
\end{align}
where the $\gamma_n \equiv \Re(s_n)<0$ and $\Omega_n \equiv |\Im(s_n)|$ are 
the damping rates and central frequencies of the first-order QNMs, respectively.
Clearly, the first and second terms in square brackets are
the ``sum tone'' and ``difference tone'' of the first-order QNMs, 
meaning that $\phi^{(2)}_{2Q}$ is {\it a superposition of the second-order QNMs}.
That is, Eq.~\eqref{eq:phi2Q_inf2} proves that the second-order QNMs are indeed excited during the evolution of the nonlinear field.
It is also clear that $C_{n{n'}}$ and $D_{n{n'}}$ represent the excitation coefficients 
of the second-order QNMs.

Next, we examine the behavior of $\phi^{(2)}_T$. Using Eq.~\eqref{eq:T_asympt}, 
we find that this term approximately behaves as {\it a power-law tail},
\beq
\phi^{(2)}_T(t,x) \approx H\biggl(\frac{1}{t-x+2} - \frac{1}{t+x+2}\biggr),
\label{eq:tail}
\eeq 
where
\beq
H \equiv \sum_{n,n'}\biggl( \frac{C_nC_{n'}}{s_n+s_{n'}}+\frac{C_nC^*_{n'}}{s_n+s^*_{n'}}+\cc \biggr)
\label{eq:H}
\eeq
is the excitation amplitude of the tail.
Equation \eqref{eq:tail} implies that this tail behaves as $t^{-1}$ for $x< t \alt 2x$, 
and as $ t^{-2}$ for $t \agt 2x$ (note that we have assumed that $x\gg1$, but not assumed that $t \gg x$).

Interestingly, this power-law tail is {\it essentially different} 
from the tails known in first-order perturbation theory \cite{Ching95}.
 First, the tails in the first-order theory originate
from the branch-cut part  $G_B$  of the Green's function (which has been neglected in our analysis),
whereas our power-law tail appears from the flat part $G_F$. 
Second, the first-order tails decay more quickly than $t^{-2}$ \cite{Ching95},
while our tail decays {\it more slowly than} $t^{-2}$.
Third, the behavior of the first-order tails depends on the form of $V(x)$,
but that of our tail is {\it independent of} it.
In fact, as we will see later, 
it is determined by the long-range behavior of the source term $S^{(2)}(x)$, or $F(x)$.
This fact implies that our tail is generated by an {\it essentially nonlinear} mechanism.  
To distinguish our tail from those in the first-order theory, 
we call it the {\it second-order power-law tail}.

In the above analysis, we have not computed the contribution from the QNM part of the Green's function, 
or $\phi_Q^{(2)}$. 
One would want to evaluate this by substituting the approximate form of $G_Q$
[Eq.~\eqref{eq:G_Q}] into Eq.~\eqref{eq:phi2_sol}. 
Unfortunately, such a calculation would not lead to a correct answer. 
The problem is that the dominant contribution to $\phi_Q^{(2)}$ comes from $x\approx0$, 
where the asymptotic approximation becomes unreliable. 
Nevertheless, we can expect, from the physical meaning of $G_Q$,
that $\phi^{(2)}_Q$ will be a linear combination of the {\it first-order} QNMs $(\propto e^{s_n (t-x)})$.
In fact, as shown in the next section, 
the first-order QNMs appear not only in $\phi^{(1)}$, but also in $\phi^{(2)}$. 
It is noted, however, that first-order QNMs appearing in $\phi^{(2)}$ are not important,
since they behave the same as the first-order perturbation $\phi^{(1)}$.    
On the contrary, the second-order QNMs $\phi^{(2)}_{2Q}$ and power-law tail $\phi^{(2)}_T$ {\it are} important, because they behave differently from $\phi^{(1)}$ .

It will be helpful to mention the origins of the second-order QNMs and tail. 
In the first-order perturbation theory, QNMs are produced 
by the QNM part $G_Q$ of the Green's function,
and power-law tails by the branch-cut part $G_B$.
This means that they are both generated {\it in curved regions}:
 the first-order QNMs are excited around the peak of the potential barrier 
and the first-order tails are yielded by the long-range decay of the potential.
On the other hand, the second-order QNMs and power-law tail arise from the ``flat part'' $G_F$. 
This means that they are generated {\it in asymptotically flat regions}.
In particular, the second-order power-law tail comes from the ``edge'' of the source distribution,
or the truncation of the first-order QNMs, at $u=0$ [see Eq.~\eqref{eq:phi2T} and Fig.~\ref{fig:spacetime}(c)].
Therefore, the second-order tail cannot appear in frequency-domain analysis,
 where the truncation is not taken into account.
 
With this fact, we can easily understand why a nonoscillating tail
 can be produced by oscillating QNMs.
As illustrated in Fig.~\ref{fig:spacetime}(c),
an observer at $(t,x)$ ``sees'' the edge extending from $(u',v')=(0,t-x)$ to $(u',v')=(0,t+x)$ 
(the observer could not ``clearly'' see the edge at $v'<t-x$,
because waves generated in this region are partly scattered
 by the potential barrier before arriving at the observer).
Since the first-order perturbation is constant along this edge,
waves generated there also have a constant phase and hence {\it do not oscillate}.

It is also easy to see that the form of the second-order tail is completely determined 
by the  spatial dependence of the source term, i.e., $F(x)$. 
In fact, the second-order perturbation generated at the edge $u'=0$ can be simply calculated to be 
\begin{align}
-\frac{1}{4} & \int_0^u du' \int_{t-x}^{t+x}dv'\phi^{(1)}(t',x')^2F(x')\delta(u') \notag \\
& \propto \int_{t-x}^{t+x}dv' F(v'/2) \propto \frac{1}{t-x+2} - \frac{1}{t+x+2},
\label{eq:tail_prop}
\end{align}
where we have used that $\phi^{(1)} = {\rm constant}$ and $x'=t'=v'/2$ along the edge, and that $F(x') = (|x'|+1)^{-2}$.
It is clear from Eq.~\eqref{eq:tail_prop} that the second-order power-law tail in Eq.~\eqref{eq:tail} is essentially the integral of $F$.

We summarize the main results of our analytic calculations. 
We have proven that the second-order QNMs do appear in the solution to the second-order perturbation.
Also, we have found that the excitation of the second-order QNMs is
accompanied by a slowly decaying power-law tail.
This ``second-order power-law tail'' is essentially different from 
those known in the first-order theory.
In particular, its behavior is determined by the source term, not by the potential term.
This means that the second-order tail is of nonlinear origin.

\section{Numerical simulations}
The above analysis has been largely dependent on two approximations:
the perturbative approximation up to second-order and the asymptotic approximation used to construct the Green's function.
In this section, we numerically confirm that these approximations are actually valid for an appropriate initial condition. 

\subsection{Setup and method}
We performed two types of simulations.
In the first type, we integrated the first- and second-order perturbation equations,
\eqref{eq:wave1} and \eqref{eq:wave2}.
In the second type, we integrate the full-order equation \eqref{eq:wave0}.
For both types of simulations, we use the same initial condition, potential term, and source term.

As an initial condition, we chose a momentarily stationary Gaussian wave packet 
\begin{gather}
f(x) = \exp[-(2.0x)^2], \qquad g(x) = 0.
\end{gather}

As the potential term $V(x)$, we adopted the P\"{o}schl-Teller potential
\beq
V_{\rm PT}(x) \equiv \frac{V_0}{\cosh^2 (Kx)},
\eeq
where $V_0$ and $K^{-1}$ are the height and the width of the potential barrier, respectively. 
It is noted that the P\"{o}schl-Teller potential damps exponentially at both $x\to\pm\infty$, while the effective potential for Schwarzschild black holes [$V_\ell(r^*)$ in Eqs.~\eqref{eq:Sch1} and~\eqref{eq:Sch2}]
decays as $r_*^{-2}$ at $r_*\to+\infty$.
However, this difference does not affect the essential features of the second-order QNMs and power-law tail, since the flat part $G_F$ of the Green's function remains unchanged as long as the potential asymptotically vanishes far away from its peak (\cite{Andersson97,note2}; see also Appendix A).

The QN frequencies for $V_{\rm PT}$ are analytically known to be \cite{Ferrari84}
\beq
s_n = - K\Bigl(n+\frac{1}{2}\Bigr) -i \Bigl(V_0-\frac{K^2}{4}\Bigr)^{1/2} .
\label{eq:QNfreq}
\eeq
In the simulations, we chose the values $(V_0,K)=(5.0,1.0)$;
the ``frequency'' $s_0$ of the least-damped ($n=0$) first-order QNM
 is calculated from Eq.~\eqref{eq:QNfreq} to be 
\beq
s_0 = -0.50 -2.18i,
\label{eq:s0}
\eeq
or, in a more familiar notation, $\omega_0 = is_0 =2.18 - 0.50i$.
It follows that the frequencies of the least-damped second-order QNMs are 
$s_0+s_0 = 2s_0=2\gamma_0-2i\Omega_0$ for the ``sum tone'' mode, and $s_0+s^*_0 = 2\gamma_0$
for the ``difference tone'' mode.
We use these values to compare the numerical results with the analytic results obtained in the last section.
We do not use the overtone ($n\geq1$) modes, since their contributions are usually smaller than that of the least-damped modes and since comparison would become much more difficult.

In integrating the wave equations, we used a standard second-order finite difference scheme. 
The computational domain was taken to be $0\leq t \leq 50.0$ and $-60.0\leq x \leq 60.0$,
so that the wave fronts of the evolved perturbations never reach the spatial boundaries.
The mesh size $(\Delta t, \Delta x)$ was set to $(0.01,0.1)$, so that the Courant number 
$\nu\equiv \Delta t/\Delta x$ satisfies the stability condition $\nu\leq1$.

\subsection{Result 1: The first-order and second-order perturbations}
\begin{figure}
\centering
\includegraphics[width=8.0cm]{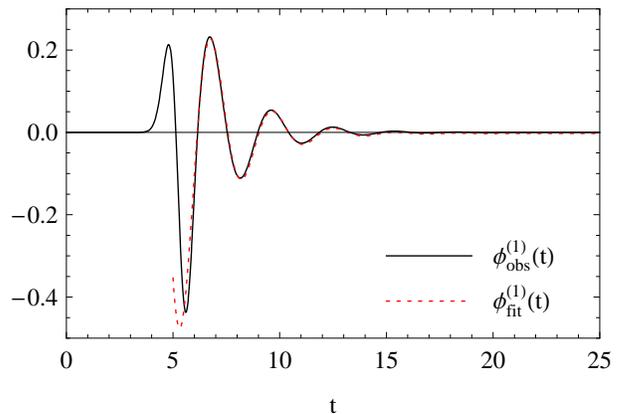}
\caption{Comparison of the numerical waveform of the first-order perturbation, $\phi^{(1)}_{\rm obs}(t)$ (solid line), with the best-fitting function for the least-damped first-order QNM, $\phi^{(1)}_{\rm fit}(t)$
 [Eq.~\eqref{eq:phi1_fit}] (dotted line). The least-damped QN frequency is $s_0=-0.50-2.18i$ [see Eq.~\eqref{eq:s0}]. The observer's position is $x_{\rm obs}=5.0$. }
\label{fig:wave1}
\end{figure}

In Fig.~\ref{fig:wave1}, we show the numerical waveform $\phi^{(1)}_{\rm obs}(t) \equiv \phi^{(1)}(t, x_{\rm obs})$ of the first-order perturbation observed at $x=x_{\rm obs} \equiv 5.0$. 
We note that its wave front arrives at $t\approx 4.0 \approx x_{\rm obs}-1.0$, not just at $t=x_{\rm obs}$.
This is because the initial wave had a width of $\delta x \sim 1$.  
From the result of Sec.~III, $\phi^{(1)}_{\rm obs}$(t) should be a superposition of the first-order QNMs. 
To evaluate the excitation coefficient $C_0$ of the least-damped mode, 
we adopt a method used in Ref.~\cite{OkuzumiSakagami07}: first, we calculate a quantity
\beq
E^{(1)} \equiv \int_{t_1}^{t_2}\Bigl[\phi^{(1)}_{\rm obs}(t)- \phi^{(1)}_{\rm fit}(t) \Bigr]^2 dt, 
\eeq
where $t_1$ and $t_2$ are fixed lower and upper limits of integration, and $\phi^{(1)}_{\rm fit}(t)$ is a fitting function defined by [cf. Eq.~\eqref{eq:phi1_asympt}]
\begin{align}
\phi^{(1)}_{\rm fit}(t) &\equiv C_0 e^{s_0(t-x_{\rm obs})} + C^*_0 e^{s^*_0(t-x_{\rm obs})} \notag \\
&= 2|C_0|e^{\gamma_0(t-x_{\rm obs})}\cos[\Omega_0(t-x_{\rm obs})-\arg(C_0)],
\label{eq:phi1_fit}
\end{align}
where $C_0\equiv|C_0|e^{i\arg(C_0)}$ is the excitation coefficient of the least-damped first-order QNM,
and $s_0$ is the frequency of the least-damped first-order QNM for the P\"{o}schl-Teller potential [Eq.~\eqref{eq:s0}].
The fitting parameter is $C_0$ (complex).
Then, we search for the value of $C_0 \equiv C_{0,{\rm fit}}$ that minimizes $E^{(1)}$.
The value of $t_1$ is taken to be larger than $x_{\rm obs}$, because the overtone QNMs will not be negligible at very early times. 
We chose $t_1=6.0$ and $t_2=25.0$, obtaining $C_{0,{\rm fit}} = -0.176-0.221i$. 
For comparison, we plotted in Fig.~\ref{fig:wave1} the function $\phi^{(1)}_{\rm fit}(t)$ with the best-fitting parameter $C_0=C_{0,{\rm fit}}$. It is clear that the least-damped mode dominates the numerical waveform except for very early times $4.0\alt t \alt 5.5$, for which the overtone modes will not be negligible.

\begin{figure}
\centering
\includegraphics[width=8.0cm]{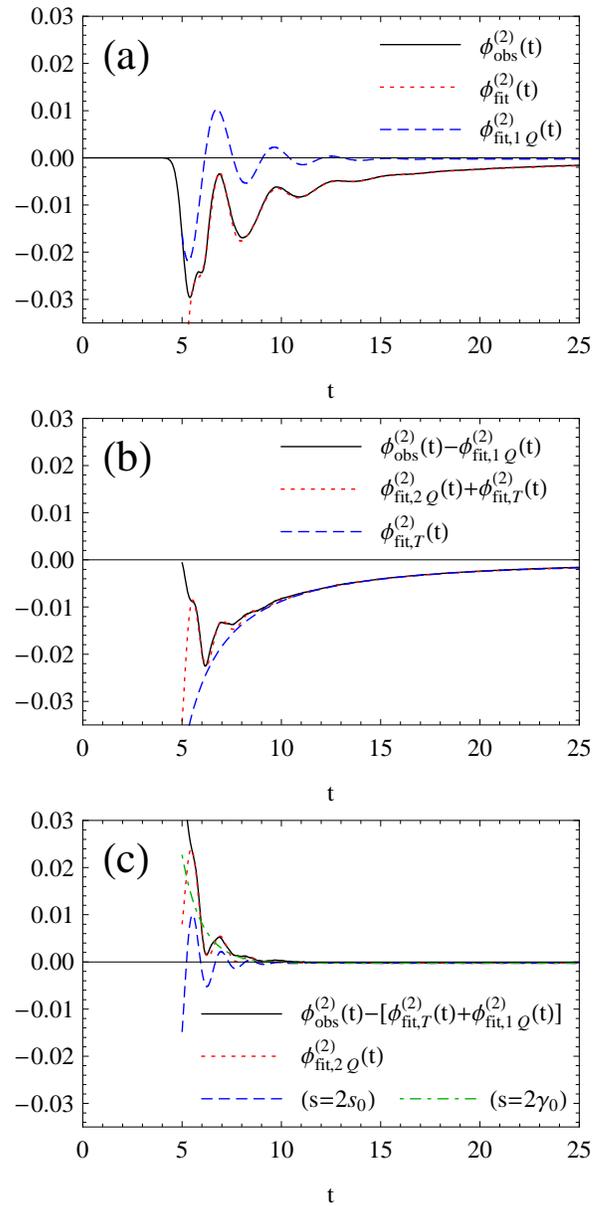}
\caption{Comparison of the numerical waveform and the best-fitting functions for the second-order perturbation $\phi^{(2)}$ at $x_{\rm obs}=5.0$. 
(a) The numerical waveform $\phi^{(2)}_{\rm obs}(t)$ (solid) and the best-fitting function for the least-damped first-order QNM $\phi^{(2)}_{{\rm fit},1Q}(t)$ [Eq.~\eqref{eq:phi2_fit1Q}] (dotted line).
(b) The numerical waveform with the first-order QNM removed, $\phi^{(2)}_{\rm obs}(t)-\phi^{(2)}_{{\rm fit},1Q}(t)$ (solid line), compared with the sum of the best-fitting functions for the second-order QNMs and power-law tail, $\phi^{(2)}_{{\rm fit},2Q}(t)+\phi^{(2)}_{{\rm fit},T}(t)$ [Eqs.~\eqref{eq:phi2_fit2Q} and~\eqref{eq:phi2_fitT}] (dotted line), and the best-fitting function for the second-order power-law tail only, $\phi^{(2)}_{{\rm fit},T}(t)$ (dashed line).
(c) The numerical waveform with all the components other than the second-order QNMs removed,
$\phi^{(2)}_{\rm obs}(t)-[\phi^{(2)}_{{\rm fit},T}(t)+\phi^{(2)}_{{\rm fit},1Q}(t)]$ (solid line), and the best-fitting function for the second-order least-damped QNMs, $\phi^{(2)}_{{\rm fit},2Q}(t)$  (dotted line), which is a superposition of the ``sum tone'' mode $\propto e^{2s_0 t}$ (dashed line) and ``difference tone'' mode $\propto e^{2\gamma_0 t}$ (dot-dashed line), where $s_0=\gamma_0-i\Omega_0$ is the frequency of the least-damped first-order QNM [see Eq.~\eqref{eq:s0}].
 }
\label{fig:wave2}
\end{figure}

In Fig.~\ref{fig:wave2}(a), we show the observed waveform of the second-order perturbation,
$\phi^{(2)}_{\rm obs}(t) \equiv \phi^{(2)}(t,x_{\rm obs})$. 
From the result of Sec. III,
we expect that $\phi^{(2)}_{\rm obs}(t)$ should be composed of three types of components:
the second-order QNMs, second-order power-law tail, and first-order QNMs. 
To decompose $\phi^{(2)}_{\rm obs}(t)$  into these components, 
we introduce a fitting function $\phi^{(2)}_{\rm fit}(t)$ defined by [cf. Eqs.~\eqref{eq:phi2Q_inf} and~\eqref{eq:tail}]
\begin{align}
\phi^{(2)}_{\rm fit}(t) 
&\equiv 
\phi^{(2)}_{{\rm fit},2Q}(t) + \phi^{(2)}_{{\rm fit},T}(t) + \phi^{(2)}_{{\rm fit},1Q}(t),
\end{align}
\begin{align}
\phi^{(2)}_{{\rm fit},2Q}(t) &\equiv \alpha_{2Q} C_{00,{\rm fit}}e^{2s_0(t-x_{\rm obs})} \notag\\
&\qquad\quad + |\alpha_{2Q}|D_{00,{\rm fit}}e^{2\gamma_0(t-x_{\rm obs})} +\cc , \label{eq:phi2_fit2Q}\\
\phi^{(2)}_{{\rm fit},T}(t) &\equiv \alpha_TH_{\rm fit}
\Bigl(\frac{1}{t-x_{\rm obs}+2}-\frac{1}{t+x_{\rm obs}+2}\Bigr),
\label{eq:phi2_fitT} \\
\phi^{(2)}_{{\rm fit},1Q}(t) &\equiv C^{(2)}_0 e^{s_0(t-x_{\rm obs})} + C^{(2)*}_0 e^{s^*_0(t-x_{\rm obs})} + \cc,
\label{eq:phi2_fit1Q}
\end{align}
where $C_{00,{\rm fit}}$, $D_{00,{\rm fit}}$, and $H_{\rm fit}$ are the excitation coefficients of the second-order QNMs and the power-law tail defined by Eqs.~\eqref{eq:C}, \eqref{eq:D}, and \eqref{eq:H}
with $C_0 = C_{0,{\rm fit}}$ and $C_n = 0$ for $n\geq1$.
The fitting parameters are $C^{(2)}_0$ (complex) 
and two scaling parameters $\alpha_{2Q}$ (complex) and $\alpha_T$ (real). 
The scaling parameters $\alpha_{2Q}$ and $\alpha_{T}$ are introduced to evaluate 
how accurately the asymptotic approximation predicts the amplitude of the second-order components.
Searching for a set of the best-fitting parameters $(\alpha_{2Q},\alpha_{T},C^{(2)}_0)_{\rm fit}$ as done for $\phi_{\rm obs}^{(1)}$, we obtain $\alpha_{2Q,{\rm fit}}=-0.0277+0.211i$, $\alpha_{T,{\rm fit}} = 0.528$, and $C^{(2)}_{0,{\rm fit}} = -0.00799-0.0102i$. We find an excellent agreement between the best-fitting function $\phi^{(2)}_{\rm fit}(t)$ and the numerical waveform $\phi^{(2)}_{\rm obs}(t)$, as shown in Fig.~\ref{fig:wave2}(a). 
The fact that $|\alpha_{2Q,{\rm fit}}|,\alpha_{T,{\rm fit}}<1$ implies that the asymptotic approximation overestimates the true amplitude of the evolved perturbation by a factor.
This is because the asymptotic approximation breaks down near the potential peak, $|x|,|x'|\alt 1$, where the source term $S^{(2)}$ gives a large contribution to $\phi^{(2)}$ [recall our choice of $F(x)$ and see Eq.~\eqref{eq:S}]. 
It should be noted that the asymptotic approximation {\it correctly} predicts 
what types of components construct the evolved second-order perturbation,
 which we really want to know in this study. 

As explained in Sec.~III, what we are interested in is the second-order waveform from which
the first-order QNM component is removed, $\phi_{\rm obs}^{(2)}(t)-\phi^{(2)}_{{\rm fit},1Q}(t)$.
We plot this in Fig.~\ref{fig:wave2}(b), comparing with the ``truly second-order'' components, $\phi^{(2)}_{{\rm fit},2Q}(t) + \phi^{(2)}_{{\rm fit},T}(t)$.
We find that the tail component dominates the extracted waveform at all times, but nevertheless the second-order QNM component is not negligible at $t\alt 9$. 
In Fig.~\ref{fig:wave2}(c), we plot the numerical waveform with all the components other than the second-order components removed, $\phi^{(2)}_{\rm obs}(t)-[\phi^{(2)}_{{\rm fit},T}(t)+\phi^{(2)}_{{\rm fit},1Q}(t)]$.
We find that the extracted numerical waveform can be well reproduced by a superposition $\phi^{(2)}_{{\rm fit},2Q}(t)$ of the least-damped second-order QNMs (i.e., sum tone $\propto e^{2s_0 t}$ and difference tone $\propto e^{2\gamma_0 t}$).

\subsection{Result 2: The fully nonlinear field}
\begin{figure}[t]
\centering
\includegraphics[width=8.0cm]{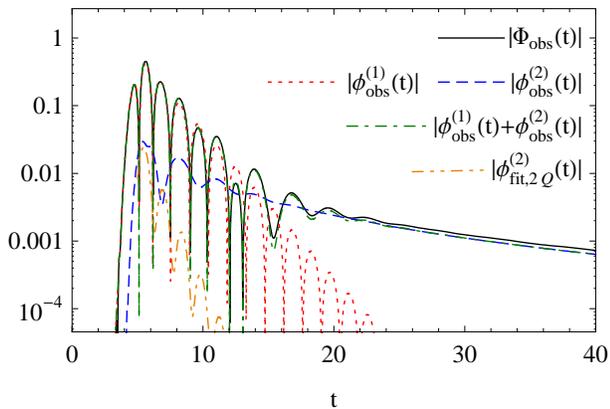}
\caption{Numerical waveforms of the fully nonlinear field (solid line), the first-order perturbation (dotted line), and the second-order perturbation (dashed line). 
Also plotted are the first-order perturbation with the second-order correction added (dot-dashed line) and
the best-fitting function for the second-order least-damped QNMs obtained in Sec.~IV.B [Eq.~\eqref{eq:phi2_fit2Q}] (dot-dot-dashed line).}
\label{fig:wavefull}
\end{figure}
In Fig.~\ref{fig:wavefull}, we compare the result of the fully nonlinear calculation of Eq.~\eqref{eq:wave0}, $\Phi_{\rm obs}(t) \equiv \Phi(t,x_{\rm obs})$, with the numerical waveform of the first- and second-order perturbations, $\phi^{(1)}_{\rm obs}(t)$  and $\phi^{(2)}_{\rm obs}(t)$.
At early times, $t\alt12$, no significant difference is seen 
between $\Phi_{\rm obs}(t)$ and $\phi^{(1)}_{\rm obs}(t)$.
This means that the first-order theory is sufficiently accurate and 
thus any higher-order correction is unimportant in this stage.
At late times $t\agt12$, however, $\phi^{(1)}_{\rm obs}$ 
fails to predict the behavior of $\Phi_{\rm obs}$.
Surprisingly, it is the second-order perturbation $\phi^{(2)}_{\rm obs}$ that shows a good agreement
with $\Phi_{\rm obs}$ in this stage.
This means that the second-order perturbation dominates the late-time behavior of the nonlinear waveform,
and {\it the perturbations of all the other orders are not significant}.
We have also confirmed that $\phi^{(2)}_{\rm obs}$ continues to agree with $\Phi_{\rm obs}$
as late as $t=50$, which strongly suggests that third and higher-order perturbations 
never surpass the second-order one in amplitude for all late times.

\section{Conclusion and Discussions}
In this study, we have investigated the nonlinear evolution of black hole ringdown
using second- and higher-order perturbation theory and a simplified nonlinear field model
for black hole metric perturbations.
We have proven that second-order QNMs, whose existence has been predicted by recent works~\cite{IokaNakano07,NakanoIoka07}, do appear in the evolved second-order perturbation.
As a bonus, we have discovered that the second-order QNMs are accompanied by a new type of power-law tail.
This power-law tail, to which we have referred as the second-order power-law tail, decays more slowly than 
any of the tails in the first-order theory, and even dominates the fully nonlinear evolution at late times.
In other words, the first-order (i.e., linear) perturbation theory 
{\it fails} to predict the late-time evolution of the ringdown, 
for which higher-order corrections {\it must} be taken into account.    
Also, we have shown that the behavior of the second-order tail is 
determined by the asymptotic form of an effective source term, i.e.,
a term expressing the nonlinearity of the system. 
Since the asymptotic form of the source term in our model is
 the same as that in black hole perturbations, 
the second-order power-law tail as well as the second-order QNMs
will certainly appear in real black hole ringdown.
These nonlinear components could open a new precision science in gravitational-wave studies.

Interestingly, the surprising failure of linear perturbation theory has been
reported just recently for other nonlinear fields
 \cite{Bizon07PRD,Bizon07CQG,Szpak07a,Szpak07b}.
Power-law tails of nonlinear origin have been first 
discovered in the Skyrme model \cite{Bizon07PRD},
 and later in a spherically symmetric Yang-Mills field 
on flat and Schwarzschild backgrounds \cite{Bizon07CQG}
(in fact, their nonlinear tails are of third-order perturbations, 
because the second-order ones cancel out;
as mentioned in Ref.~\cite{Bizon07CQG}, these seem to be special cases). 
In both cases, the nonlinear tails decay more slowly than the linear ones, 
and do dominate the fully nonlinear evolution at late times.
From this fact, we suspect that our second-order power-law tail would 
have close relation to these nonlinear tails.
However, we emphasize that the nonlinear tail {\it of metric perturbations} 
has been for the first time discovered in this study.

In our future works, we plan to address the following open questions:
\begin{itemize}
\item[(i)] {\it Do the second-order QNMs and power-law tail appear in observable quantities, i.e., in gravitational waves?}
As explained in Sec.~II, the perturbations $\phi^{(1)}$ and $\phi^{(2)}$ of our model field correspond,
respectively, to the master functions $\psi^{(1)}_{\ell m}$ and $\chi^{(2)}_{\ell m}$ of the perturbations of the Schwarzschild geometry [see Eqs.~\eqref{eq:Sch1},~\eqref{eq:Sch2},~\eqref{eq:wave1}, and~\eqref{eq:wave2}].  
Therefore, it is straightforward to apply our analysis used in this study to the evolution of the master functions, although the computations will be much more complicated.
It should be noted, however, that these master functions themselves are not observable quantities.
In order to convert them into truly observable quantities, we have to perform a gauge transformation
from the Regge-Wheeler (RW) gauge to the asymptotically flat (AF) gauge, and then compute the transverse-traceless (TT) parts of the metric perturbations in the AF gauge \cite{IokaNakano07,NakanoIoka07}.
We will have to prove that the second-order QNMs 
and power-law tail remain to appear in the TT parts of the metric perturbation. 
Giving a full-nonlinear proof should be a nice challenge for numerical relativity.

\item[(ii)] {\it Do our results remain unchanged under more realistic initial conditions?}
In the present study, we have assumed initial perturbations localized around the potential peak.
In order to deal with astrophysically important problems, 
we need to extend our analysis to more general initial conditions.
One of the most tractable problems among them will be the head-on collision 
of equal-mass Schwarzschild black holes \cite{PricePullin94,Anninos95}. 
The evolution of the second-order perturbations during the collision was already studied 
\cite{Gleiser96PRL} using Misner's initial data \cite{Misner60}, 
but the second-order QNMs and power-law tail were not reported in that study. 
Therefore, we must discuss this issue using the same initial data.
We expect that the effect of initial perturbations located far away from the potential peak
would not be significant,
since the effective source term is suppressed  at $r\gg2M$ by a factor of $\propto r^{-2}$. 

\item[(iii)] {\it Is it possible to detect the second-order power-law tail (if it exists) 
with future gravitational-wave detectors?}
To address this issue, we need to estimate the energy carried out by the second-order tail assuming astrophysically realistic events (e.g., binary black hole mergers), 
as already done for second-order QNMs \cite{IokaNakano07,NakanoIoka07}.
Also, we have to develop a data analysis to extract the second-order tail from detector outputs.
While the matched filtering technique has been developed for extracting QNMs \cite{Tsunesada05},
there seems to be only a few techniques for power-law tails \cite{Blanchet95}.
If the second-order tail is detectable, it could be used to distinguish true signals and spurious ones in black hole ringdown search, where fake reduction and event identification are crucial \cite{NakanoIoka07,Tsunesada05}.
\end{itemize}
\begin{acknowledgments}
We are grateful to H. Nakano for fruitful discussions.
We thank W. Varela for a careful reading of the manuscript.
This work is supported in part 
by the Grant-in-Aid from the 
Ministry of Education, Culture, Sports, Science and Technology
(MEXT) of Japan, Nos.18740147 and 19047004 (K.I.)
\end{acknowledgments}

\appendix

\section{Construction of Green's function}
In this appendix, we summarize the derivation of the flat part $G_F$ and QNM part $G_Q$ of the Green's function using the asymptotic approximation \cite{Andersson97}.

First, we take the Laplace transform of $G(\tau;x,x')$ with respect to $\tau$:
\beq
\tilde{G}(s;x,x') \equiv \int_0^{\infty}d\tau\,e^{-s\tau}G(\tau;x,x'). 
\eeq
This is an analytic function of $s$ defined for $\Re(s)>0$.
Taking the Laplace transform of Eq.~\eqref{eq:waveG} as well, we obtain 
\beq
\Bigl[ \pd_x^2 -s^2 - V(x) \Bigr]\tilde{G}(s;x,x') = \delta(x-x').
\label{eq:Gs}
\eeq
To construct the solution $\tilde{G}$, we introduce two
homogeneous solutions to Eq.~\eqref{eq:Gs}, $y_L(s,x)$ and $y_R(s,r)$, which satisfy the boundary conditions
\begin{align}
y_L(s,x) &\approx \begin{cases}
e^{sx}, & (x\to-\infty) \\
A_L(s)e^{sx} + B_L(s)e^{-sx}, & (x\to+\infty)
\end{cases}
\label{eq:y_L} \\
y_R(s,x) &\approx \begin{cases}
A_R(s)e^{-sx} + B_R(s)e^{sx}, & (x\to-\infty) \\
e^{-sx}, & (x\to+\infty)
\label{eq:y_R}
\end{cases}
\end{align}
where $A_L,\,B_L,\,A_R,\,$ and $B_R$ are functions of $s$ determined by the form of the potential barrier.
Physically, $|B_{L(R)}(s)/A_{L(R)}(s)|$ and $|1/A_{L(R)}(s)|$ mean, respectively, the reflection amplitude and the transmission amplitude of a wave incident from $x\to+\infty(-\infty)$ with unit amplitude and a frequency $\omega=is$.
With these homogeneous solutions, $\tilde{G}(s;x,x')$ is constructed to be 
\beq
\tilde{G}(s;x,x') = \begin{cases}
\dfrac{1}{W(s)}y_L(s,x)y_R(s,x'), & (x<x') \\
\dfrac{1}{W(s)}y_L(s,x')y_R(s,x), & (x'<x)
\end{cases}
\label{eq:Gs_sol}
\eeq
where $W(s)$ is the Wronskian defined by
\beq
W(s) \equiv y_L(s,x)\pd_x y_R(s,x) - y_R(s,x)\pd_x y_L(s,x).
\label{eq:W}
\eeq
It can be easily shown that $W(s)$ is independent of $x$. Evaluating the right-hand side of Eq.~\eqref{eq:W} at $x\to\pm\infty$ gives $W(s) = -2sA_L(s) = -2sA_R(s)$. Equation~\eqref{eq:Gs_sol} is the only
possible choice of $\tilde{G}$ for the boundedness of time-domain solutions \cite{NollertSchmidt92,Nollert99}.
\begin{figure}
\centering
\includegraphics[width=6cm]{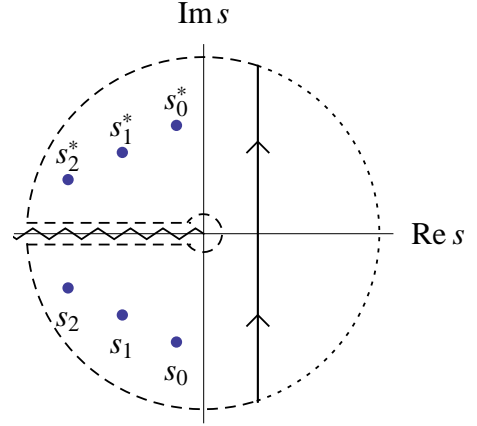}
\caption{Schematic illustration of the contour for the integration in Eq.~\eqref{eq:G_inverseL}. The solid line represents the original contour. The zigzag line is a (possible) branch cut for the integrand.}
\label{fig:contour}
\end{figure}

Taking the inverse Laplace transform of $\tilde{G}$, we have
\begin{align}
G(\tau;x,x') &= \int_{\epsilon-i\infty}^{\epsilon+i\infty}\frac{ds}{2\pi i}\,e^{s\tau}\tilde{G}(s;x,x')\qquad \notag \\
&= -\int_{\epsilon-i\infty}^{\epsilon+i\infty}\frac{ds\,e^{s\tau}}{4\pi isA_L(s)} \notag \\
&\quad\times
\begin{cases}
y_L(s,x)y_R(s,x'),& \quad(x<x')\\
y_L(s,x')y_R(s,x),& \quad(x'<x)
\end{cases}
\label{eq:G_inverseL}
\end{align}
where $\epsilon>0$.
This integral can be evaluated by closing the contour of integration with an infinite semicircle in either the left or right half in the complex $s$-plane (if $\tilde{G}$ has a branch point at $s=0$, which is true in many important cases, the infinite path is taken to circumvent the branch cut; see Fig.~\ref{fig:contour}). For this purpose, we analytically continue the integrand in Eq.~\eqref{eq:G_inverseL} into the left half of the complex $s$-plane, $\Re(s)<0$.
The analytic continuation of the integrand has an infinite number of poles $\{s_n\}$, which can be shown to
be equal to the zeros of $A_L(s)$. This implies that analytic continuation of the homogeneous solution $y_L(s,x)$ satisfies the ``outgoing boundary conditions'' \eqref{eq:QNM} at $s=s_n$, meaning that $y_n(x)\equiv y_L(s_n,x)$ are the first-order QNMs. 

It is generally hopeless to carry out the integration in Eq.~\eqref{eq:G_inverseL} exactly since neither $y_L$, $y_R$, nor $A_L$ can be written as a simple function of $s$. 
Nevertheless, it is possible to carry out the integration approximately by using the fact that $y_L$ and $y_R$
approach to exponential forms far away from the potential, $|x|\gg1$ [see Eqs.~\eqref{eq:y_L} and \eqref{eq:y_R}]. 
We call this ``asymptotic approximation'' after Andersson \cite{Andersson97}.
From Eqs.~\eqref{eq:y_L} and \eqref{eq:y_R}, Eq.~\eqref{eq:G_inverseL} is approximately written as
\onecolumngrid
\begin{align}
G(\tau;x,x') &\approx -\int_{\epsilon-i\infty}^{\epsilon+i\infty}\frac{ds}{4\pi is} \times \begin{cases}
 \dfrac{1}{A_L(s)}e^{s(\tau-(x-x'))}, & (x'\ll -1) \\
  e^{s(\tau-|x-x'|)} + \dfrac{B_L(s)}{A_L(s)}e^{s(\tau-(x+x'))}, & (x'\gg 1) 
\end{cases}
\label{eq:G_asympt}
\end{align}
where we have assumed that $x\gg1$ for all the cases.

Now we choose the contour of integration according to the convergence of the integrand in Eq.~\eqref{eq:G_asympt} at $|s|\to\infty$. 
For $x'\ll-1$, the integrand converges in the left half-plane if $\tau-(x-x')>0$,
and in the right half-plane otherwise.
Taking a contour in the half-plane where the integrand converges and applying the residue theorem, we obtain
\beq
G(\tau;x,x') \approx \begin{cases}
-\dfrac{1}{2}\displaystyle\sum_{n=0}^{\infty}\dfrac{e^{s_n(\tau-x+x')}}{s_nA_L'(s_n)}, & (\tau>x-x')\\
0, & (\tau<x-x') 
\label{eq:G_-}
\end{cases}
\eeq
where $A_L'(s_n) = dA_L(s)/ds|_{s=s_n}$.
For $x'\gg1$, the integrand involves two exponentials with different arguments, and thus a preferred choice of the contour is not obvious.
Following Ref.~\cite{Andersson97}, we choose to close the contour in the left half-plane if $\tau-(x+x')>0$, and in the left half-plane otherwise. This yields
\begin{align}
G(\tau;x,x') \approx \begin{cases}
-\dfrac{1}{2}\displaystyle\sum_{n=0}^{\infty}\dfrac{B_L(s_n)}{s_nA_L'(s_n)}e^{s_n(\tau-(x+x'))}, 
& (\tau>x+x') \\
-\dfrac{1}{2}, & (\tau<x+x'~{\rm and}~\tau>|x-x'|) \\
0. & ( \tau<|x-x'|)
\label{eq:G_+}
\end{cases}
\end{align}
Combining Eqs.~\eqref{eq:G_-} and \eqref{eq:G_+} together with Eq.~\eqref{eq:QNM} gives
$G_F$ and $G_Q$ in Eqs.~\eqref{eq:G_F} and \eqref{eq:G_Q}.
It is noted that the ``tail part'' $G_B$ can not derived from the asymptotic approximation, 
since the asymptotic approximation breaks down in the low-frequency region ($|s|\approx 0$).  

\twocolumngrid


\begin{thebibliography}{99}
\bibitem{note1}
At the horizon ($r_*\to-\infty$), $S^{(2)}_{\ell m}(t,r_*)$ actually behaves as $\propto e^{r*/2M}$, 
not as $\propto r_*^{-2}$. 
However, this difference does not affect $\phi^{(2)}_F$ (i.e., second-order QNMs and tail),
since $G_F(t-t';x,x')$ vanishes for $x'<0$ assuming an observer's position at $x>0$.
\bibitem{note2}
In contrast, the ``branch-cut part'' $G_B$ changes in response to the behavior of the potential far away from its peak.  As a consequence, the P\"{o}schl-Teller potential does not yield any power-law tail in the perturbations, while the potential of Schwarzschild black holes does yield a (first-order) tail decaying as $t^{-2\ell-2}$. However, this tail is not significant compared to the second-order tail because the second-order one decays as $t^{-2}$, i.e., more slowly than the first-order one.
\end{thebibliography}
\end{document}